\documentclass[]{spie}  %>>> use for US letter paper
\usepackage[]{graphicx}
\usepackage[usenames, dvipsnames]{color}
\usepackage[colorlinks=true, allcolors=blue]{hyperref}

\title{The Infrared Imaging Spectrograph (IRIS) for TMT: advancing the data reduction system} 

\author{
Gregory L. Walth\supit{a},
Shelley A. Wright\supit{a,b},
Nils-Erik Rundquist\supit{a,b},
David Andersen\supit{c},
Edward Chapin\supit{c},
Eric Chisholm\supit{d},
Tuan Do\supit{e},
Jennifer Dunn\supit{c},
Brent Ellerbroek\supit{d},
Kim Gillies\supit{d},
Yutaka Hayano\supit{f},
Chris Johnson\supit{e},
James Larkin\supit{e},
Takashi Nakamoto\supit{f},
Reed Riddle\supit{g},
Luc Simard\supit{c},
Roger Smith\supit{g},
Ryuji Suzuki\supit{f},
Ji Man Sohn\supit{e},
Robert Weber\supit{g},
Jason Weiss\supit{d},
Kai Zhang\supit{h}
\skiplinehalf
\supit{a} Center for Astrophysics \& Space Sciences, University of California San Diego, CA, 92039, USA; \\
\supit{b} Department of Physics, University of California San Diego, CA, 92039, USA; \\
\supit{c} National Research Council of Canada - Herzberg, Victoria, BC, V9E2E7 Canada; \\
\supit{d} Thirty Meter Telescope Observatory Corporation, Pasadena, CA 91105 USA; \\
\supit{e} Physics \& Astronomy Department, University of California Los Angeles, CA 90095 USA; \\
\supit{f} National Astronomical Observatory of Japan, Osawa, Mitaka, Tokyo, 181-8588 Japan; \\
\supit{g} Caltech Optical Observatories, 1200 E California Blvd., Pasadena, CA 91125 USA; \\
\supit{h} National Astronomical Observatories / Nanjing Institute of Astronomical Optics \& Technology, Chinese Academy of Sciences, Nanjing 210042, China \\
}

\authorinfo{Further author information: (Send correspondence to G.L.W.)\\G.L.W..: E-mail: gwalth@ucsd.edu}
\begin{document} 
\maketitle 

%%%%%%%%%%%%%%%%%%%%%%%%%%%%%%%%%%%%%%%%%%%%%%%%%%%%%%%%%%%%% 
\begin{abstract}
Infrared Imaging Spectrograph (IRIS) is the first light instrument for the Thirty Meter Telescope (TMT) that consists of a near-infrared (0.84 to 2.4 micron) imager and integral field spectrograph (IFS) which operates at the diffraction-limit utilizing the Narrow-Field Infrared Adaptive Optics System (NFIRAOS). The imager will have a 34 arcsec x 34 arcsec field of view with 4 milliarcsecond (mas) pixels.  The IFS consists of a lenslet array and slicer, enabling four plate scales from 4 mas to 50 mas, multiple gratings and filters, which in turn will operate hundreds of individual modes.  IRIS, operating in concert with NFIRAOS will pose many challenges for the data reduction system (DRS).  Here we present the updated design of the real-time and post-processing DRS. The DRS will support two modes of operation of IRIS: (1) writing the raw readouts sent from the detectors and performing the sampling on all of the readouts for a given exposure to create a raw science frame; and (2) reduction of data from the imager, lenslet array and slicer IFS. IRIS is planning to save the raw readouts for a given exposure to enable sophisticated processing capabilities to the end users, such as the ability to remove individual poor seeing readouts to improve signal-to-noise, or from advanced knowledge of the point spread function (PSF). The readout processor (ROP) is a key part of the IRIS DRS design for writing and sampling of the raw readouts into a raw science frame, which will be passed to the TMT data archive. We discuss the use of sub-arrays on the imager detectors for saturation/persistence mitigation, on-detector guide windows, and fast readout science cases ($<$ 1 second).  
\end{abstract}

%>>>> Include a list of keywords after the abstract 

\keywords{integral field spectroscopy, data reduction pipeline}

%%%%%%%%%%%%%%%%%%%%%%%%%%%%%%%%%%%%%%%%%%%%%%%%%%%%%%%%%%%%%
\section{INTRODUCTION}\label{sec:intro}  % \label{} allows 

IRIS is the first light instrument for TMT, operating at the diffraction-limit with a near-infrared (0.84 to 2.4 micron) imager and IFS. The imager will have four 4k$\times$4k Hawaii-4RG detectors with a 34 arcsec$\times$34 arcsec field of view (FoV) using a 4 mas pixel scale. The IFS will have one Hawaii-4RG which will have four spatial scales (lenslet:4 mas, 9 mas; slicer: 25 mas, 50 mas).  The IFS will have 12 gratings with resolutions R=4000, 8000, and 10000. With the sequential optical design, both the imager and IFS will use the same set of 68 filters. There will be hundreds of configurations that IRIS will support.

The next generation of large telescopes will provide unique challenges for data reduction pipelines. IRIS is designed to operate its imager and IFS simultaneously, and therefore the DRS needs to support hundreds of modes with real-time reduction for acquisition and analysis. The DRS will handle both lenslet and slicer IFS data supporting all resolutions. Another key feature of the DRS is that it communicates with TMT, NFIRAOS, and IRIS, storing the relevant metadata into each of the individual readout FITS headers. 

%Walth et al.\cite{Walth2016} presented the preliminary design of the DRS.

This paper describes the current design for the IRIS DRS and some of the instrument and telescope design challenges. Section 2 gives an overview of the DRS. Section 3 describes the DRS data rates and storage requirements. Section 4 discusses the On-Instrument Wavefront Sensors potential vignetting of IRIS and how the DRS will mitigate it. Section 5 discusses the imager sub-array use cases and the DRS plans to handle them.  Finally, Section 6 discuses the potential use of the Atmospheric Dispersion Corrector (ADC) to implement a low dispersion mode on the imager.

\begin{figure}[ht!]
\begin{center}
\caption{\label{fig:drs_flow}DRS flow chart showing its interactions with the instrument sequencer, the detector HCDs, the TMT DMS and the storage disks.}
\includegraphics[width=.8\linewidth]{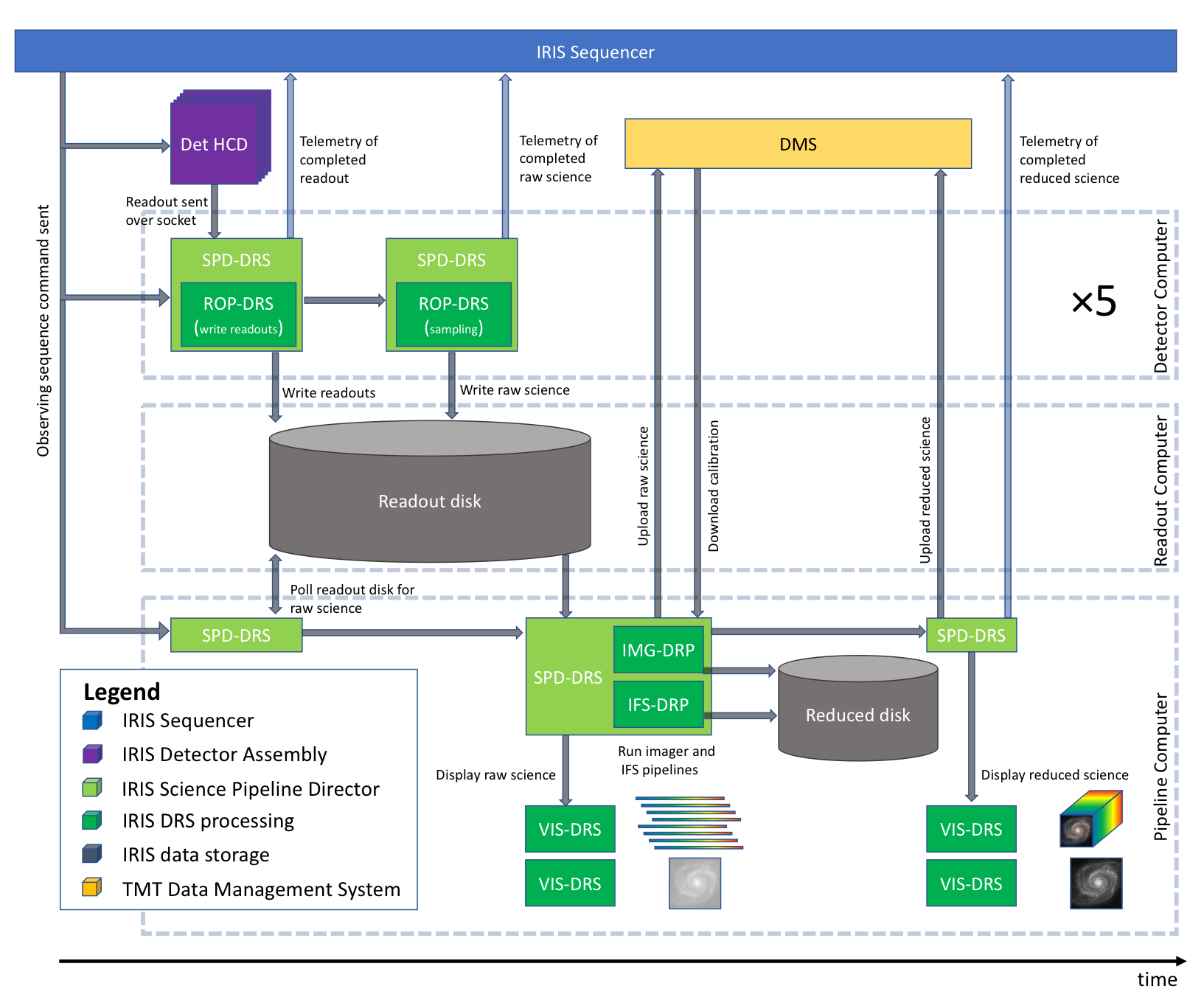}
\end{center}

\end{figure}

\section{DRS Overview}

The IRIS DRS will play a crucial role in handling the detector processing, post-processing reduction, as well as visualization and analysis of the data. The initial design of the IRIS DRS is described in Ref. \citenum{Walth2016}. Here we describe the updated design from the preliminary design review (PDR). Figure \ref{fig:drs_flow} shows each of the interactions and timings between the DRS, IRIS instrument sequencer, detectors, and TMT data management system (DMS).

We briefly describe the interactions and timings shown in Figure \ref{fig:drs_flow}. First, the IRIS instrument sequencer sends an event which is received by the detectors and the science pipeline director (SPD-DRS) indicating that an observing sequence is about to occur. The SPD-DRS communicates with the readout processor (ROP-DRS) to prepare to receive and process a detector data sequence. When individual detector frames are read out, they open a socket to the ROP-DRS that then writes them to the readout disk. After all of the frames from an exposure are read out, the ROP-DRS performs the user selected sampling [e.g., Up-the-ramp (UTR)] to create and store a single raw science frame on the readout disk. Telemetry needed for real-time processing for each raw science frame is collected by the SPD-DRS from the TMT DMS, which includes meta-data for each data sequence from the IRIS instrument sequencer; executive software (ESW); adaptive optics executive software (AOESW); telescope control system (TCS); and NFIRAOS Science Calibration Unit (NSCU). The SPD-DRS then uploads the the raw science frames to the TMT DMS.

Once raw science frames are written, the SPD-DRS will use the required calibration files (e.g., flats, rectmat, darks) from the TMT DMS. After all of the necessary calibrations files are downloaded (including sky frames for the IFS), the real-time imager (IMG-DRP) and IFS (IFS-DRP) pipelines are run on the raw science frames. The reduced science frames are stored on a reduced storage disk and uploaded to the TMT DMS for archiving. The SPD-DRS will also send the raw and reduced data to the visualization software (VIS-DRS) for real-time inspection and analysis.

\subsection{Brief description of terminology}
\label{sec:term}
\begin{figure}[ht]
\begin{center}
\caption{\label{fig:readouts}Example observing sequence of readouts and ramps which are part of an individual raw science frame.}
\includegraphics[width=.8\linewidth]{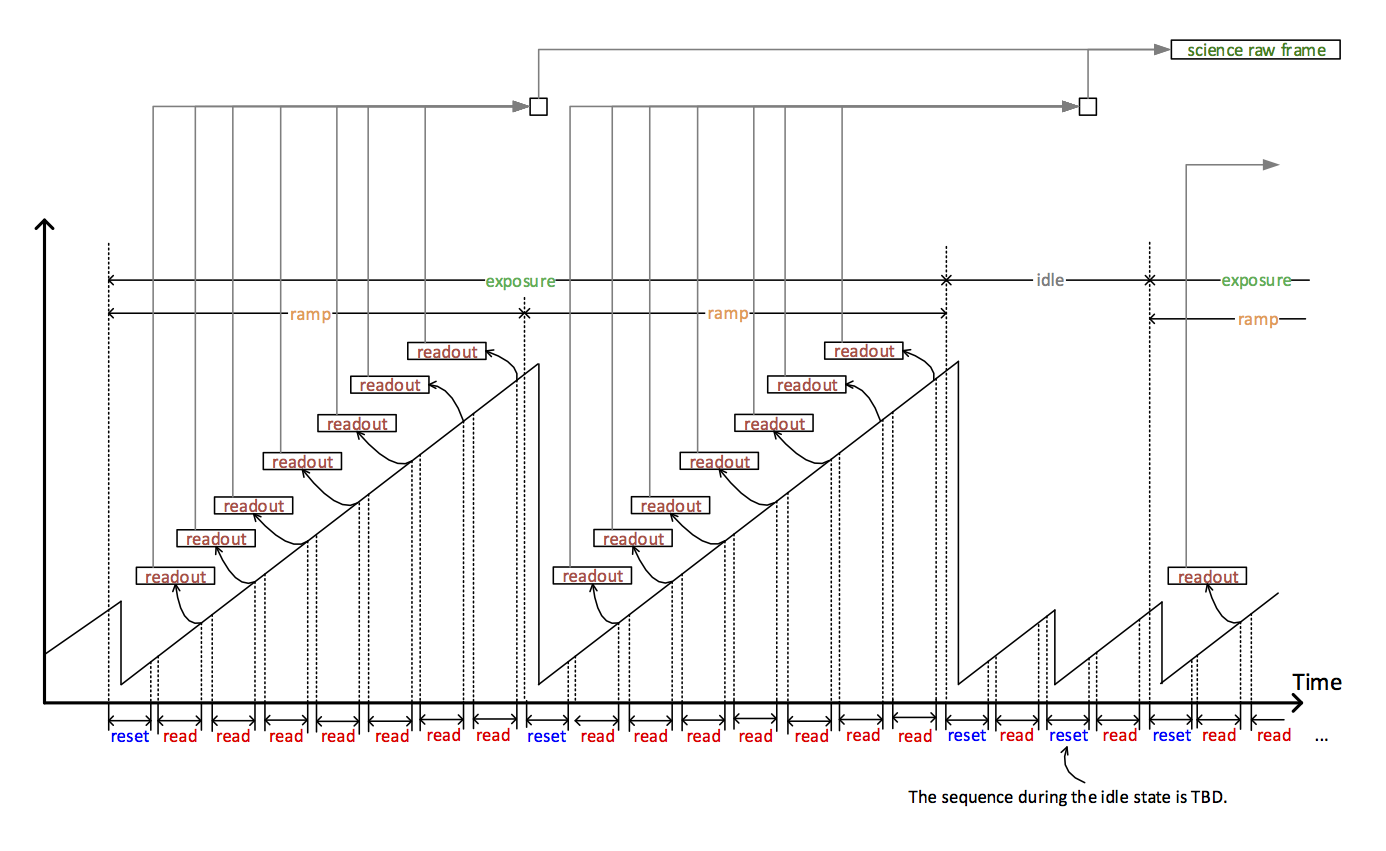}
\end{center}
\end{figure}

We describe the terminology used for the data reduction system and the readout processing and sampling as illustrated in Figure \ref{fig:readouts}:
\begin{itemize}
\item Exposure: All of the readouts combined using the selected sampling mode.
\item Ramp: All of the readouts between the resets.  Depending on the sampling mode this would involve the configured number of co-adds.  This may be more than 1 set of coadds.
\item Readout: The individual read from the detector.
\item Raw science frame: The same as an individual exposure.
\item Reduced science frame: A single raw science frame that has been reduced.
\item Observing sequence: A series of exposures (or raw science frames).
\end{itemize}

\subsection{Science Pipeline Director}
\label{sec:spd}

The science pipeline director (SPD-DRS) is the interface layer that communicates between the instrument sequencer and reduction pipelines. The SPD-DRS receives commands (observing events) and telemetry from the instrument sequencer and then sends its status back. The SPD-DRS has several primary functions: (1) running the readout processor (ROP-DRS); (2) running the reduction pipelines; (3) displaying the raw and reduced data to the user for evaluation and analysis; and (4) uploading raw and reduced science frames to the TMT DMS. The SPD-DRS has four critical modes in operation.

\begin{enumerate}
\item The SPD-DRS instructs the readout processor (ROP-DRS) to receive and write the data to the readout disk when an observing event is received. The SPD-DRS will also instruct the ROP-DRS to sample the individual readouts into a raw science frame once the readouts start being written to disk.

\item The SPD-DRS will instruct imager real-time (IMG-DRP) and spectrograph real-time (IFS-DRP) pipelines to reduce the raw science frames when all of the necessary files are ready for reduction (i.e., raw science frames, calibration frames and sky frames). The SPD-DRS polls the readout and reduction disks for new files, by creating and updating a database of raw science frames, calibration frames and sky frames to be used during the reduction.

\item The SPD-DRS will instruct the quicklook visualization (VIS-DRS) to display raw science frames and reduced frames when they are finished processing.

\item The SPD-DRS is also responsible for uploading raw science frames and reduced frames to the TMT DMS (TBD). The individual readouts will remain stored on the readout disk.
There will be 6 concurrently running SPD-DRS processes, each running on their own computer: one for each detector assembly computer (5 computers; 1 IFS and 4 imager computers), which will receive and write the readouts (as well as sample the readouts into the raw science frame); and one on the DRS pipeline computer (which will reduce the data).
\end{enumerate}

\subsection{Readout processor}
\label{sec:rop}

The readout processor (ROP-DRS) will have two functions: (1) receiving and writing data (including receiving telemetry and creating headers) to a FITS file from the imager/IFS HCD; and (2) sampling of the individual readouts to the final generated raw science frame. Both of these processes will run simultaneously. The ROP-DRS is also part of the Data Reduction Pipeline User Package (USER-DRS) which is the offline version of the pipeline that users will need in order to process the individual readouts.

The SPD-DRS will receive an observation event from the instrument sequencer with the instrument configuration, observation configuration, and sampling method. The SPD-DRS instructs the ROP-DRS to receive the detector data over a socket [e.g., User Datagram Protocol (UDP)]. The ROP-DRS will receive the imager/IFS data through a socket and then collect all of the metadata from the subscribed telemetry to generate a header and then write an individual readout FITS file.

The ROP-DRS will perform the sampling algorithms [e.g., UTR, Multiple Correlated Double Sampling (MCDS); selected by the astronomer and passed through the instrument sequencer] on the individual readouts, which will conduct the following functions: reference pixel subtraction, linearization, bad pixel flagging; UTR/MCDS computed for each ramps; and coaddition of the ramps.

\subsection{Algorithms}
\label{sec:alg}
The final imager/IFS pipeline (F-DRP) performs the full reduction of the imager and IFS data. This pipeline utilizes the full set of algorithms for the imager and IFS and is part of the USER-DRS package. The imager real-time pipeline (IMG-DRP) performs the reduction of the IRIS imager data, which uses a subset of routines of the imager F-DRP.   Similarly, the IFS real-time pipeline (IFS-DRP) performs the reduction of the IRIS lenslet and slicer data, which uses a subset of the IFS F-DRP. The SPD-DRS instructs the real-time imager and IFS pipelines to run. Table \ref{tab:alg} shows the algorithms used for each of the pipelines.

In Ref.~\citenum{Walth2016} we outline the algorithms required by the real-time and offline pipelines. We briefly described the functionality of each of the algorithms for the real-time and off-line post-processing. The general reduction algorithms are the following:
\begin{itemize}
\item {\bf Generate master dark} - generates master dark frame from the median of 5--10 dark calibration frames.
\item {\bf Dark subtraction} - subtract dark current from the detector in the science and calibration frames using a single dark frame in the real-time, or a master dark in the final pipeline, at the same exposure time as the science and calibration frames.
\item {\bf Remove detector artifacts} - mask and/or remove hot and dead pixels from the detector in the science and calibration frames.
\item {\bf Scaled sky-subtraction} - subtract the sky-background either from the imager, which is a constant value, or the IFS, which a sky frame scaled to match the science frame.
\item {\bf Flux calibration} - calibrate science frames (imager and IFS), which are in data number per second (DN/s), to flux density erg/s/cm$^2$/Hz for the imager and erg/s/cm$^2$/Ang for the IFS by using a standard star calibration.
\item {\bf Mosaic/Combine science} - combine science frames from imager and IFS by shifting the frames based on their WCS coordinates (or dither offsets) and either median or averaging the overlapping regions.
\end{itemize}

The reduction algorithms used only by the IFS slicer and lenslet are the following:
\begin{itemize}
\item {\bf Spectral extraction} - extract spectra from the IFS. The slicer requires a spectral trace of each spectrum. The lenslet requires a deconvolution of the entire lenslet array to determine the flux for a given lenslet.
\item {\bf Wavelength calibration} - shift and resample (linearize) an individual extracted spectrum using a calibration arc lamp frame, containing typical lines found the in the near-IR, such as Ar, Kr, and Xe.
\item {\bf Cube assembly} - assemble data cube with the extracted spectra from the spectral extraction routine and map them to an x, y position on the sky (spatial rectification) based on the WCS information, and their z positions, which are shifted based on their individual wavelength solutions.
\item {\bf Residual ADC} - if needed, correct the residual from the ADC, using a lookup table.  Tests on-sky using a star at different airmasses will be needed to calibrate the lookup table.
\item {\bf Telluric correction} - remove atmospheric telluric absorption features by using a telluric standard, which is typically a featureless star.
\end{itemize}

The reduction algorithms used only by the imager are the following:
\begin{itemize}
\item {\bf Field distortion correction} - correct imager science frames for the distortion caused by the camera optics and resample to a regular grid. 
\end{itemize}

The reduction algorithms used only by the imager and IFS slicer are the following:
\begin{itemize}
\item {\bf Flat fielding} - remove variations in inter-pixel responsivity by dividing by a normalized flat calibration frame
\end{itemize}

Advanced algorithms:
\begin{itemize}
\item {\bf PSF-reconstruction} - perform deconvolution on the imager and IFS data using a simulated PSF from the NFIRAOS PSF simulator to reconstruct the PSF during an observation. The PSF-reconstructed (PSF-R) FITS files are generated by TMT Observatory Operations Software system by making use of real-time telemetry from the observatory, NFIRAOS, and IRIS. The IRIS DRS will be designed to handle these PSF-R calibration files with an associated observing sequence (e.g., integration time of imager and/or IFS). The DRS deconvolution algorithms are still being determined and are considered an advanced goal of the system.
\item {\bf Optimizing readouts} - Allowing the end user extra control of the readouts by enabling access to the readout processor, including the various sampling techniques (e.g., UTR, MCDS). This gives them the ability to remove ``bad" or poor seeing readouts.
\end{itemize}

\begin{table}[ht]
\caption{Algorithms used for the imager and IFS real-time and final data pipelines. Note: flat-fielding is slicer only}
\label{tab:alg}
\begin{center}
\begin{tabular}{|l|c|c|c|c|}
\hline
Algorithms & \multicolumn{2}{|c|}{Real-time Pipeline} & \multicolumn{2}{|c|}{Final Pipeline (F-DRP)} \\
\cline{2-5} & Imager & IFS & Imager & IFS \\
\hline
Generate master dark              & x & x & x & x \\
Dark subtraction                  & x & x & x & x \\
Correction of detector artifacts  & x & x & x & x \\
Remove cosmic rays                & x & x & x & x \\
Flat fielding*                    & x & x & x & x \\
Spectral extraction               &   & x &   & x \\
Wavelength calibration            &   & x &   & x \\
Cube assembly                     &   & x &   & x \\
Scaled sky-subtraction            & x & x & x & x \\
Residual ADC                      &   &   &   & x \\
Telluric correction               &   &   &   & x \\
Field distortion correction       &   &   & x &   \\
Flux calibration                  &   &   & x & x \\
Mosaic/Combine science                &   &   & x & x \\
\hline
Advanced algorithms & & & & \\
\hline
PSF-reconstruction                &   &   & x & x \\
Optimizing readouts               &   &   & x & x \\
\hline
\end{tabular}
\end{center}
\end{table}

\section{Data Rates and Storage}
IRIS will aim to store all of the individual readouts and therefore it is critical to characterize the total anticipated storage needed to accommodate this requirement. Ref. \citenum{Walth2016} found that using Correlated Double Sampling (CDS) with a clock time of 2.16 seconds/exposure with a 4k$\times$4k detector would generate 64 MB/exposure. This means that for 91 total observing nights per year would generate 1.8 PB. More details about this data rate can be found in Ref. \citenum{Walth2016}. In the following subsections, we extend this data rate estimate to the data products and calibrations that would need to be stored in the TMT DMS. Additionally, we investigate the maximum computing requirements for the DRS, assuming non-optimized algorithms for memory storage and processing.

\subsection{Reduced Data Rates}
We make the following assumptions to determine the reduced data rates: the reduced frames contain 3 extensions (science, noise and flags); single floats (32 bits) are used for the science and noise frames; unsigned integers are used for the flags (8 bits); IFS slicer (45x90) at R=8000 (800 spectral elements) and imager (4096x4096); each raw IFS frame has data cube created; each raw imager frame is sky-subtracted; 60 second dither overhead; and a 5 second frame overhead. We also define a few ``real world use cases" for operation. 

For our real-world case, we assume a dither pattern of 300 second IFS observations on a single object with simultaneous CDS imager observations with as many frames as possible. This assumes 16 hours observing with 12 hours on sky, 4 hours for calibration and 91 nights a year (1/4 year). \footnote{We do not combine frames for final stacks in this calculation or include calibrations (i.e., arcs, flats, master darks, master flats).}

Table \ref{tab:data_product} shows the total data rates from the reduced products for the the imager and IFS.  In addition, IRIS will have 120 modes for the IFS (slicer and lenslet), which will require a rectification matrix for each one.  Including 3 extensions in each file will require 5.6 GB/year. In general, the frame size for the imager is 604.0 MB (4 detectors x 151.0 MB) and for the IFS it is 73.7 MB (which is the maximum size for a spectrum with 3 extensions).

\begin{table}[ht]
\caption{Real world IRIS data rates from reduced products.}
\label{tab:data_product}
\begin{center}
\begin{tabular}{|l|c|c|c|c|}
\hline
Products	    & Individual Frame & Hourly Data Rate & Nightly Data Rate & Yearly Data Rate \\
\hline
Reduced Imager	& 151.0 MB  (1 det) &  62.4 GB/hr  (1 det) &  0.75 TB/night  (1 det) &  68.1 TB/year  (1 det) \\
                & 604.0 MB  (4 det) & 250.0 GB/hr  (4 det) &  3.00 TB/night  (4 det) & 272.6 TB/year  (4 det) \\
Reduced IFS  	& 58.3 MB           &	641.3 MB/hr	       & 11.0 GB/night	         & 1.0 TB/year \\
\hline
\end{tabular}
\end{center}
\end{table}

\subsection{Calibration Data Rates from the TMT DMS}
Assuming that we store the calibration data on the TMT DMS, we compute the rate at which we expect to download the calibration data throughout a night of observations. The calibration files are necessary for the real-time reduction to run on the IRIS pipeline computer.  It is imaginable that some calibration data are permanently stored on the DRS pipeline computers, though the exact amount and for how long has yet to be determined.  We go through this exercise to determine the maximun amount of data we might request from the TMT DMS. The rates include the cases in which the rectification matrices (for the IFS lenslet observations) are downloaded from the TMT DMS, and stored locally. The rates have a 1st run and nth run which represent frames that are needed initially but then can be reused throughout the night (i.e., master darks, bad pixel maps).  We also assume the modes are unique (i.e., no reuse of calibration frames for filter and scale).

Lenslet calibration data rates from the TMT DMS include the following: arc lamp, bad pixel map, master dark, sky spectra, rectification matrix (rectmat).  Data rates are calculated both with and without the rectmat downloaded from the TMT DMS.
Slicer calibration data rates from the TMT DMS include the following: arc lamp, bad pixel map, master dark, flat field, sky spectra.

Imager calibration data rates from the TMT DMS include the following: bad pixel map, master dark, sky images.
The maximum data rate from the TMT DMS for calibration frames in a given night, assuming 8 unique modes of imaging and spectroscopy, would be 7619.5 MB/night w/o rectmats and 10354.3 MB/night with rectmats. The calibration frame sizes are listed in Table \ref{tab:cal_data}.

The memory usage of the individual algorithms can be found in Table \ref{tab:comp_mem_ifs} for the IFS and Table \ref{tab:comp_mem_imager} for the imager.  It is assumed that the frames are completely loaded into memory; our calculations do not not account for the possibility of optimal methods of reading in sub-arrays of data as well as utilizing running sum techniques.

\begin{table}[ht]
\caption{Calibration data rates from the TMT DMS}
\label{tab:cal}
\begin{center}
\begin{tabular}{|l|l|l|r|r|}
\hline
\multicolumn{3}{|c|}{Instrument} & 1st Observation & nth Observation \\
\multicolumn{3}{|c|}{}            & [MB]   & [MB] \\
\hline
\hline
IFS    & Lenslet & Rectmat     & 1441.1 & 1273.3 \\\cline{3-5} 
	   &         & No rectmat  &  392.5 &  224.7 \\\cline{2-5} 
	   & \multicolumn{2}{|l|}{Slicer} &  543.5 &  224.7 \\
\hline
\multicolumn{3}{|l|}{Imager}   & 1275.2 &  604.0 \\
\hline
\end{tabular}
\end{center}
\end{table}

\begin{table}[ht]
\caption{Calibration frame sizes for a single detector}
\label{tab:cal_data}
\begin{center}
\begin{tabular}{|l|r|}
\hline
Frame &	Size [MB] \\
\hline
Arc lamp &	151.0 \\
Bad pixel map &	16.8 \\
Master dark	& 151.0 \\
Sky spectra	& 73.7 \\
Sky image &	151.0 \\
Rectification matrix (maximum) &	1048.6 \\
Flat (IFS slicer only) &	151.0 \\
\hline
\end{tabular}
\end{center}
\end{table}

\begin{table}[ht]
\caption{Memory used by specific DRS algorithm for the IFS}
\label{tab:comp_mem_ifs}
\begin{center}
\begin{tabular}{|l|c|c|l|}
\hline
DRS post processing 	  & Num. Frames & Memory & Function \\
\hline
Generate master dark      & ~10 & 1.51 GB           & Median combination \\
Dark subtraction          &   2 & 302 MB            & Subtraction \\
Remove detector artifacts &   2 & 302 MB            & Bad pixel and CR removal \\
Flat fielding*            &   2 & 302 MB            & Division by normalized flat field \\
Spectral extraction       &   1 & 178.5 - 1122.3 MB & Advanced spectral extraction \\
Wavelength calibration    &   2 & 147.4 MB          & Least square minimization \\
Cube assembly             &   2 & 147.4 MB          & Cube assembly \\
Scaled sky-subtraction    &   2 & 147.4 MB          & OH and continuum scaling \\
Residual ADC              &   2 & 147.4 MB          & Atm. Dispersion Correction \\
Telluric correction       &   2 & 147.4 MB          & Telluric feature removal \\
Flux calibration          &   2 & 147.4 MB          & Flux calibration \\
Mosaic/Combine science 	      &  $>$2 & N$_{\rm frames}$ x 73.7 MB & Dither shifts \\
\hline
\end{tabular}
\end{center}
\end{table}

\begin{table}[ht]
\caption{Memory used by specific DRS algorithm for the imager per detector. Note: FRS is frames.}
\label{tab:comp_mem_imager}
\begin{center}
\begin{tabular}{|l|c|c|l|}
\hline
DRS post processing & Num. Frames & Memory & Function \\
\hline
Generate master dark         & ~10  & 1.51 GB (6.04 GB)                   & Median combination \\
Dark subtraction             &  2   & 302 MB (1.21 GB)                    & Subtraction \\
Remove detector artifacts    &  2   & 302 MB (1.21 GB)                    & Bad pixel and CR removal \\
Flat fielding                &  2   & 302 MB (1.21 GB)                    & Division by normalized flat field \\
Scaled sky-subtraction       &  2   & 302 MB (1.21 GB)                    & Scale factor from sky frames \\
Field distortion correction  &  1   & 151 MB (604 MB)                     & Field distortion correction \\
Flux calibration             &  2   & 302 MB (1.21 GB)                    & Flux calibration \\
Mosaic/Combine science           &  $>$2  & N$_{\rm FRS}$ x 151 MB (N$_{\rm FRS}$ x 604 MB) & Dither shifts \\
\hline
\end{tabular}
\end{center}
\end{table}

\section{On-Instrument Wavefront Sensor Vignetting}
The current design of the On-Instrument Wavefront Sensor (OIWFS) includes three independently-positioned OIWFS which will patrol a region 2' in diameter. The OIFWS patrol regions each cover roughly 3/4 of the area of the imager.  This means that stars can be chosen that will fall on imager FoV. It is therefore necessary to understand the vignetting that can occur on the detector due to the OIWFS probe arms.  
%, diffraction and ghosting
The OIWFS probe arms, when over the imager, will produce a circular shadow in focal plane with 12" diameter.  This means that the OIWFS can block 7x10$^{6}$ pixels (or 42\% of a 4K x 4K detector).  Vignetting is a problem that can be dealt with either through observing planning, in which observers avoid choosing guide stars near or over of the imager, or by using on-detector guide windows (discussed in section \ref{sec:sub}), or finally, in DRS post processing. 

There are two vignetting cases: (1) full vignetting can usually be avoided by using the on-detector guide windows (ODGW), and (2) stars slightly outside the edge imager FoV, which can still partly vignette the imager.  Both cases can be mitigated in planning software, which is under development at TMT. The software needs make sure that observers do not partly vignette IFS FoV, or to allow the OIWFS probe to use the 5" FoV to center stars at edge, away from IRIS imager to minimize vignette area.

In the cases where vignetting is unavoidable, the DRS will generate mask of probe arms projected onto the IRIS FoV if there is partial or complete vignetting. The DRS will need engineering calibration data during the integration phase (INT), which will map the vignetting 2D profile and flat fielding across IRIS. The DRS will also require telemetry of probe arm positions which will be retained in each IRIS science raw frame headers for post-processing.  If telemetry of probe arms are within the limits of vignetting then header keywords are flagged for the DRS to handle later in post-processing. The DRS would use a mask in the individual modules (e.g., sky/dark subtraction, flat fielding, detector artifacts, mosaicking).

\section{Sub-arrays}
\label{sec:sub}
IRIS will have three modes for which sub-arrays will be used: ODGWs, sub-arrays for saturation/persistence mitigation, and sub-array science cases. The DRS will receive telemetry from the imager with the number and size of the sub-arrays for post-processing.  

With a 30-meter primary aperture, TMT/IRIS will saturate more routinely than 10-meter class telescopes.  For example, in a 1-second exposure the imager will saturate a source in J-band at 13.9 Vega and a 10 second exposure will saturate at 16.4 Vega. This poses a significant problem for most imaging programs since many of the well-known stars from 2MASS and SDSS will saturate. Science cases that require high precision astrometry will need a way to mitigate this saturation such that the stars can be useful. In addition, saturation of the detector is a major problem since these pixels are essentially lost for hours due to persistence.

Using sub-arrays is a possible way to mitigate this saturation issue. The technical feasibility of such an idea is possible with the current Hawaii-2RG detectors and could be applied to Hawaii-4RG detectors.  The current design of the readout electronics enables the read out of the sub-array.  The only limitation is the rate at which the sub-arrays are read.  

A typical operation using sub-arrays would read out the science frames at a fixed cadence, while a 4x4 sub-array will be read at 1800 Hz (600 Hz for a 6x6).  However, there are concerns about reading out at such rates, where the detector is heated at the location of the sub-array which would creating a temperature gradient. The rate of self-heating is estimated to be 800 e-/K, which could affect dark current and be seen as a glow on the detector.  During the IRIS final design phase (FDP) a study testing sub-arrays will be performed to assess the potential effects of using sub-arrays for saturation mitigation.

There are many science cases in which the current readout time of 1 second is inadequate.  In order to study variability on much shorter timescales (10 ms - 1 s)  it may be necessary to read out the detector faster, in which would be possible for a sub-arrays. A few of the fast readout science cases are blazar jets, Sgr A*, cataclysmic variables and TNO/Kuiper Belt object occultations.  For additional time variable science cases refer to Ref. \citenum{Skidmore2015}.

\subsection{Sub-arrays for saturation/persistence}

\begin{figure}[ht!]
\begin{center}
\caption{\label{fig:sat}(LEFT) Number of pixels saturated in a 1s exposure
in each of the broad band filters. Sub-array sizes of 4x4 will only mitigate up to a magnitude
brighter stars.  It may be necessary to investigate larger sub-arrays, such
as 6x6 or greater. (RIGHT) Number of pixels saturated in a 1s exposure in each of the narrowband filters.  Each line style represents a different filter in a filter set (e.g., solid, dash, dash-dotted, dotted for Zn1, Zn2, Zn3, Zn4). Sub-array sizes of 6x6 enable saturation/persistence mitigation for fields with stars 1-1.5 magnitudes brighter.  Larger sub-arrays will enable brighter magnitude limits.}
\includegraphics[width=.49\linewidth]{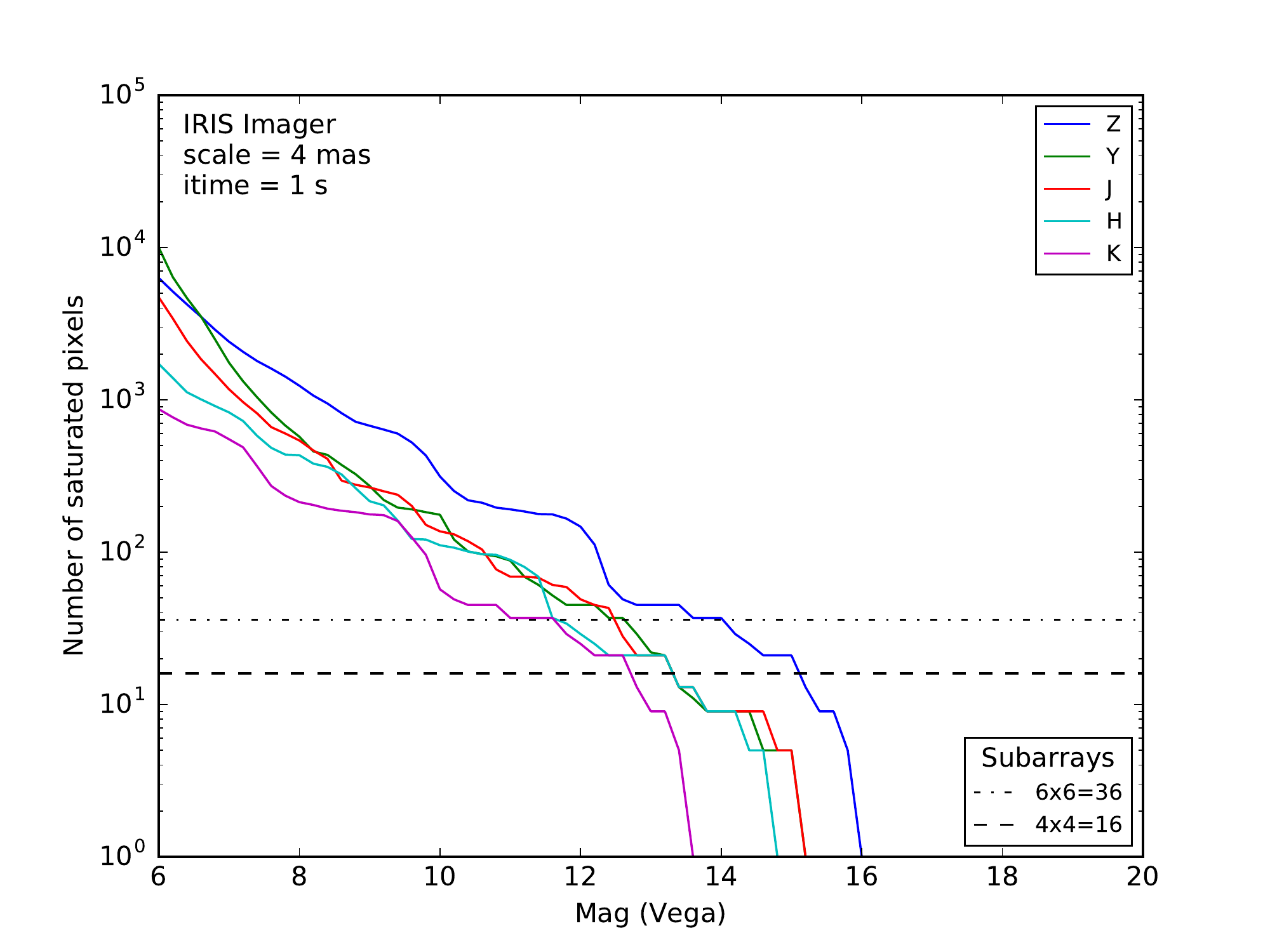}
\includegraphics[width=.49\linewidth]{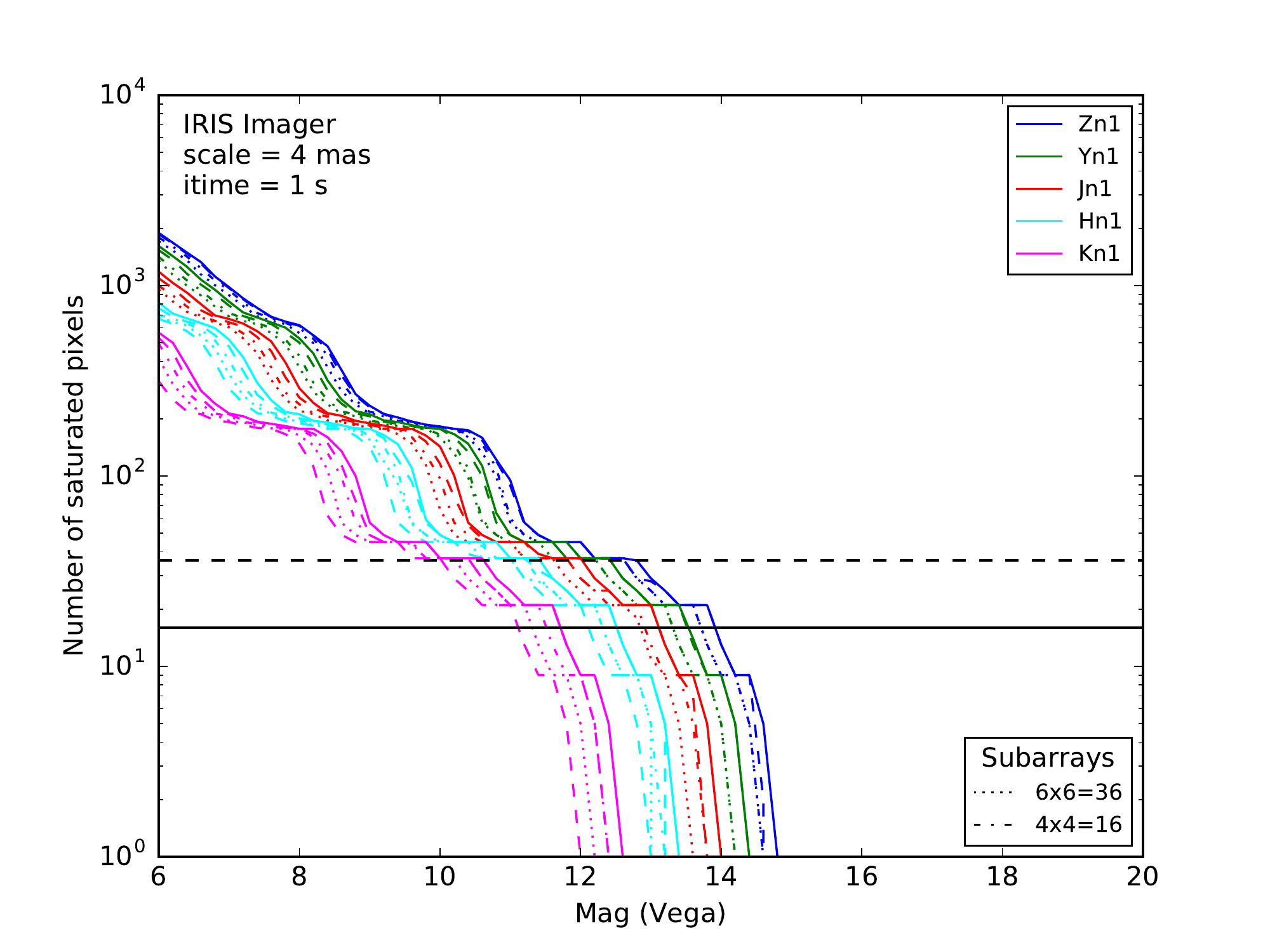}
\end{center}
\end{figure}

\begin{figure}[ht!]
\begin{center}
\caption{\label{fig:sat_zband}Simulated images of the saturated pixels for
different magnitude stars in the Z-band. A 16x16 sub-array will enable
exposures of sources  5 magnitudes brighter in saturation/persistence
mitigation.  However, the thermal effect of resetting these pixels (in the form of glow) will radiate out beyond the sub-array size and need to be masked out by the DRS.}
\includegraphics[width=.6\linewidth]{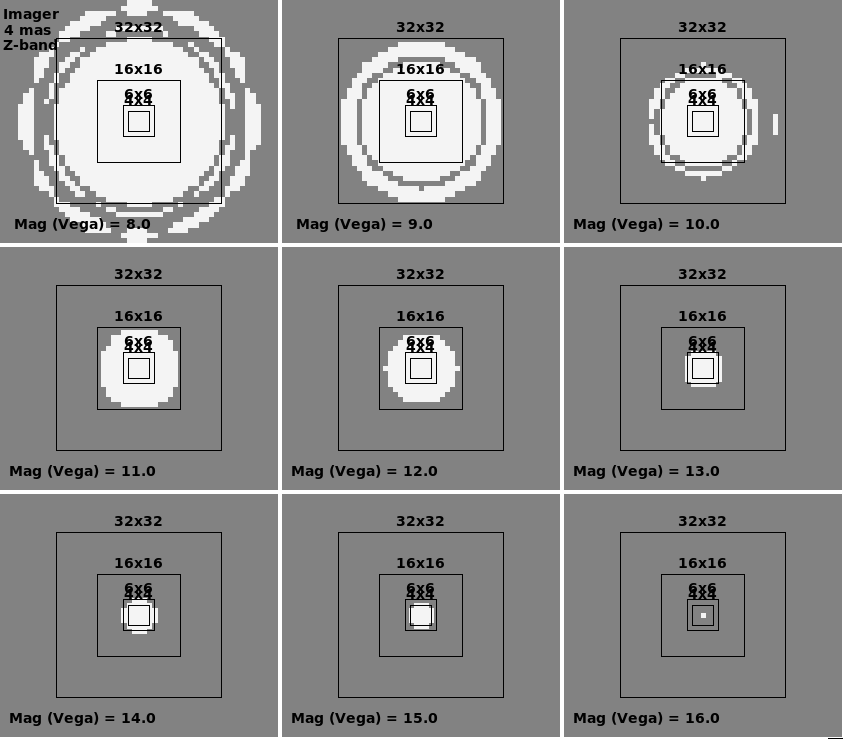}
\end{center}
\end{figure}

It is expected that saturation will affect the detectors in the form of
persistence, which could require on the order of hours to fully dissipate.
The Hawaii-2 detectors had significant issues with persistence which could
last an entire night. The Hawaii-2RG mitigated the some of the persistence
issues, seen in the previous generation of detectors.  Now for bright
sources, saturating sources can still cause significant problems, but
only in the first few exposures after saturation.  One possible solution
for dealing with persistence from saturation is to constantly read out the
sub-array that is affected by a bright star.  This requires pre-planning
such that it is known upon which pixels bright sources will fall. Another
effect is the thermal heating of pixels constantly being read out.  This
will cause ``blooming" outside of the sub-array.  Using the IRIS
simulator~\cite{Do2014, Wright2010}, various sub-array sizes were simulated
with the broad and narrow band filters (Figure \ref{fig:sat}). It has been
proposed to use 4x4 or 6x6 sub-arrays, however they would only increase the
effective dynamic range by 0.5 or 1.5 magnitudes respectively.  Larger
sizes may be needed to cover a wider range of magnitudes (Figure
\ref{fig:sat_zband}).

\subsection{On-detector guide windows}

The OIWFS is expected to vignette a significant portion of the IRIS imager (~42\% of a 4Kx4K detector). In cases which an astronomer has no choice but to pick guide stars that will fall on the detector, it is possible to use sub-arrays as on-detector guide windows (ODGW) instead of the OIWFS. However, this is an explicit trade-off between limiting potential AO performance/sky coverage and vignetting (i.e., an ODGW is not a full replacement for an OIWFS).

The ODGW sub-arrays will be sent to the OIWFS and the last readout is what will remain when the full frame is read out.  It is planned to send all of the frames of the ODGWs from the OIWFS over a socket. The current design of the ODGWs is such that only one ODGW per detector (4 total). It is anticipated that the ODGW readouts would be combined using the ROP-DRS and stitched into the raw science frames, however specific details are still under development. 

\subsection{Mitigation}

To mitigate the effects of sub-arrays, the DRS will need to generate a mask
of all sub-arrays per image. The mask will include sub-array coordinates
that are registered to individual raw science frames, and masks will need
to be included in the DRS functions (e.g., sky/dark subtraction, flat
fielding, detector artifacts, mosaicking).  During the fabrication and
integration phases, it may be necessary to develop additional DRS modules
and functions from engineering data to deal with ``blooming" and dark
current effects from the heating of detector pixels outside the sub-array.
It may be the case that this will have to be dealt with in post-processing.

\begin{figure}[ht!]
\begin{center}
\caption{\label{fig:adc}(TOP-LEFT) RGB image of the {\it Hubble} eXtreme
Deep Field (XDF) using the filters F105W, F125W and F160W. The red box
marks the region of deepest XDF coverage. The yellow box represents the
IRIS 34 arcsec$\times$34 arcsec FoV. (TOP-RIGHT) Zoom-in of the IRIS FoV of the XDF. (BOTTOM-LEFT) Preliminary simulation of the XDF in the Y-band, assuming the sources are all point sources. (BOTTOM-RIGHT) IRIS low dispersion mode simulation of XDF (with the ADC maximally dispersed).
}
\includegraphics[width=.9\linewidth]{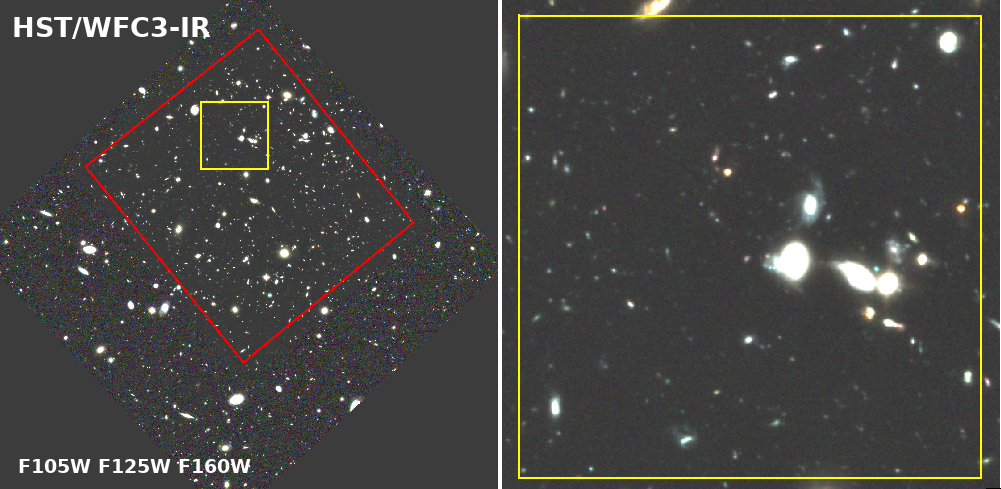}
\includegraphics[width=.42\linewidth]{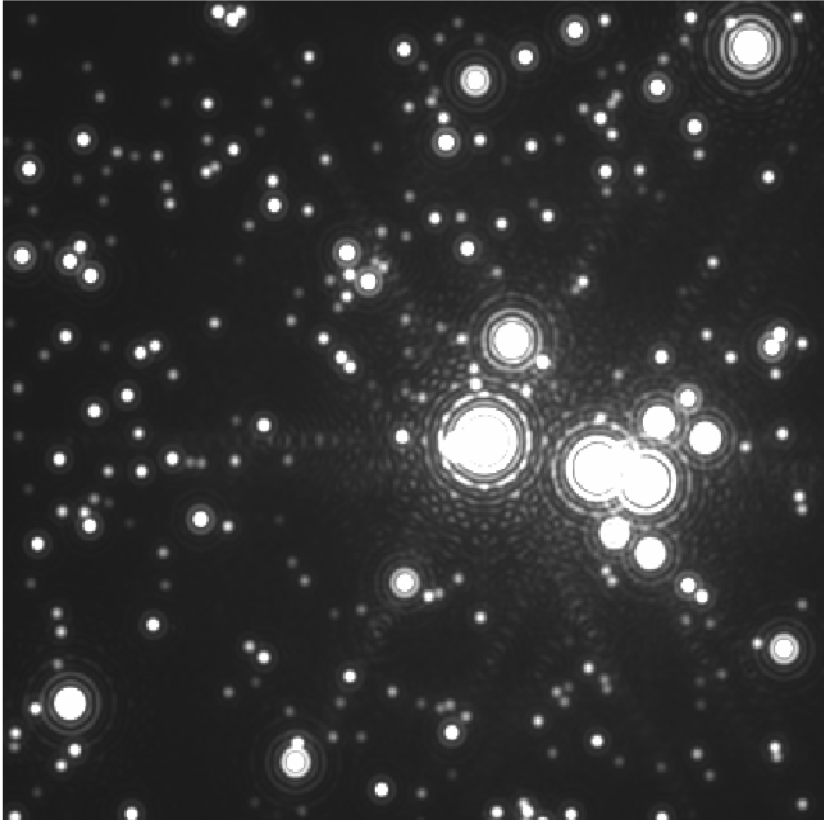}
\hspace{0.2cm}
\includegraphics[width=.42\linewidth]{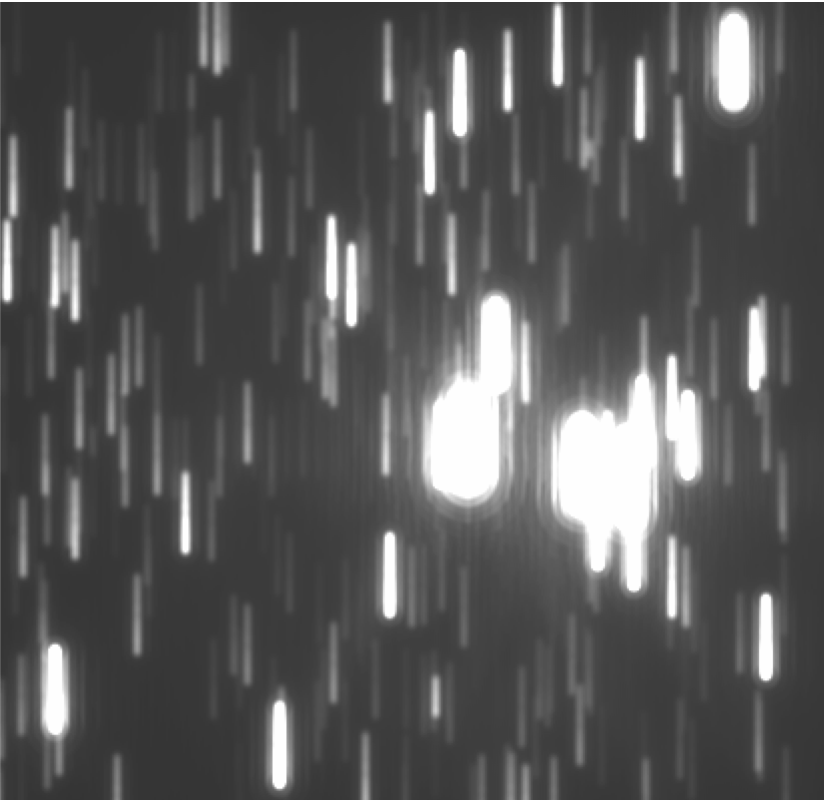}

\end{center}
\end{figure}

\section{Low Dispersion Mode}

The atmospheric dispersion corrector (ADC) corrects for the refraction
caused by the atmosphere, at varying airmass (or elevation). The use of an
ADC is important for IRIS, especially since it will be operating at the
diffraction limit of TMT (see Ref. \citenum{Phillips2016}) which is 6 - 17 mas (z-band to K-band). At 45$^{\circ}$ elevation (airmass $\sim$ 1.4), the atmospheric dispersion ranges from 100 to 10 mas (z-band to K-band) [see Ref. \citenum{Phillips2016}] which significantly impacts the resolution achievable.

A potential mode being investigated during FDP is using the ADC as a low dispersion prism (R=50-100) where the imager would act as a slitless spectrograph. For example, at zenith the images would require no ADC correction, and you could use the ADC to maximally disperse the light. At lower elevations, you would then rotate the ADC to partially compensate the increased atmospheric
dispersion, thus resulting in an approximately constant effective
dispersion throughout the observation.

% 6 - 17 mas, z-K

The low dispersion mode would present challenges to the DRS, such as: (1) the bright near-IR sky background; (2) confusion between nearby sources; (3) contrast between the brightest and faintest sources; and (4) calibration throughout the night. Typically at medium resolution (e.g., R=4000), the near-IR background is fainter between the OH sky lines, which makes possible to detect fainter sources. However, at low resolution the near-IR sky background becomes an issue as the OH skylines blend together, which approaches the imaging near-IR sky background. 

The IRIS science team is interested in exploring the functionality of this
low resolution spectroscopic mode during FDP. Questions we hope to address for this mode are: what are the possible resolutions, what is the lowest elevation you can observe, what are the sensitivity limits, and how will the DRS support this mode?  We may be able to lean on previous software developments and methodology used with HST/WFC3-IR grism\cite{Kummel2009, Brammer2012, Ryan2018}.

Figure \ref{fig:adc} shows preliminary simulations of the {\it Hubble}
eXtreme Deep Field (XDF)\cite{Illingworth2013}. We are currently developing
a module within the IRIS simulator~\cite{Wright2016, Wright2014, Do2014, Wright2010} to add the functionality of simulating science fields with the low dispersion mode to investigate observing and calibration strategy as well as sensitivity limits.

\section{Summary}
We have presented the updated DRS design for IRIS for the TMT. One of the
biggest changes to the design is that the detectors now send the individual
readouts through a socket to the readout processor.  Additionally, the
readout processor will write the readouts and perform sampling (e.g.,
UTR) on an exposure (series of reads and ramps). In this updated
design, the science pipeline director acts as the interface between the
instrument sequencer and the various DRS pipelines, as well as sending raw
and reduced images to the visualization tools. The science pipeline
director will also request all of the telemetry from the relevant TMT, IRIS
and NFIRAOS subsystems.  We also investigate the data storage and
computation requirements for running the DRS during an observing campaign,
and find that they are modest. The storage requirements only become
challenging when we want to store all of the individual readouts to the
archive. The OIWFS can vignette the IRIS imager and IFS (even slightly
outside the FoV), which can be mostly solved either with the TMT planning
tool or with ODGWs.  If the imager does get vignetted, the DRS will need to
investigate masking based on the OIWFS telemetry.  With TMT's 30-meter
aperture, saturation will be a significant issue for IRIS. One of the ways
to mitigate it is through the use of sub-arrays.  However, resetting the
detector or read out a much faster rate within the sub-array will heat the
detector. The DRS will employ masking beyond the reset regions to account
for the ``blooming" effects from the heating.  Finally, we investigate a potential new mode, utilizing the IRIS ADC as a low dispersion prism with the IRIS imager, which has the advantage of the imager throughput and the potential disadvantage of the high sky background.

%%%%%%%%%%%%%%%%%%%%%%%%%%%%%%%%%%%%%%%%%%%%%%%%%%%%%%%%%%%%%
\acknowledgments     %>>>> equivalent to \section*{ACKNOWLEDGMENTS}       
 
The TMT Project gratefully acknowledges the support of the TMT
collaborating institutions.  They are the California Institute of
Technology, the University of California, the National Astronomical
Observatory of Japan, the National Astronomical Observatories of China and
their consortium partners, the Department of Science and Technology of
India and their supported institutes, and the National Research Council of
Canada.  This work was supported as well by the Gordon and Betty Moore
Foundation, the Canada Foundation for Innovation, the Ontario Ministry of
Research and Innovation, the Natural Sciences and Engineering Research
Council of Canada, the British Columbia Knowledge Development Fund, the
Association of Canadian Universities for Research in Astronomy (ACURA) ,
the Association of Universities for Research in Astronomy (AURA), the U.S.
National Science Foundation, the National Institutes of Natural Sciences of
Japan, and the Department of Atomic Energy of India.  

%%%%%%%%%%%%%%%%%%%%%%%%%%%%%%%%%%%%%%%%%%%%%%%%%%%%%%%%%%%%%
%%%%% References %%%%%
% Cherenkov instruments (veritas,hawk, FAMOUS, cosmic house, nicholas law)
% SETI (general, optical/nir)
% structural (popko, stiffness)
% fresnel
\bibliography{report}   %>>>> bibliography data in report.bib
\bibliographystyle{spiebib}   %>>>> makes bibtex use spiebib.bst

\end{document}